# An SDN hybrid architecture for vehicular networks: Application to Intelligent Transport System

Soufian Toufga [1], Philippe Owezarski [1], Slim Abdellatif [2], Thierry villemur [3]
1 LAAS-CNRS, Université de Toulouse, CNRS, Toulouse, France
2 LAAS-CNRS, Université de Toulouse, CNRS, INSA, Toulouse, France
3 LAAS-CNRS, Université de Toulouse, CNRS, UT2J, Toulouse, France
Email:{toufga, owe, slim, villemur}@laas.fr

*Abstract*—Vehicular networks are one of the cornerstone of an Intelligent Transportation System (ITS). They are expected to provide ubiquitous network connectivity to moving vehicles while supporting various ITS services, some with very stringent requirements in terms of latency and reliability. Two vehicular networking technologies are envisioned to jointly support the full range of ITS services : DSRC (Dedicated Short Range Communication) for direct vehicle to vehicle/Road Side Units (RSU) communications and cellular technologies. To the best of our knowledge, approaches from the literature usually divide ITS services on each of these networks according to their requirements and one single network is in charge of supporting the each service. Those that consider both network technologies to offer multi-path routing, load balancing or path splitting for a better quality of experience of ITS services assume obviously separately controlled networks.

Under the umbrella of SDN (Software Defined Networking), we propose in this paper a hybrid network architecture that enables the joint control of the networks providing connectivity to multi-homed vehicles and, also, explore the opportunities brought by such an architecture. We show through some use cases, that in addition to the flexibility and fine-grained programmability brought by SDN, it opens the way towards the development of effective network control algorithms that are the key towards the successful support of ITS services and especially those with stringent QoS. We also show how these algorithms could also benefit from information related to the environment or context in which vehicles evolve (traffic density, planned trajectory, ..), which could be easily collected by data providers and made available via the cloud.

## I. INTRODUCTION

The evolution of cars technology deals with assisting the driver on the road. Cars are more and more equipped with sensors in order to cope with several driving tasks as automatic switching of the lights or wipers, warning the driver if he goes out of his driving lane, front and backward detectors for urgent breaking or backward driving assistance, park assist, etc. Going further, the next generation of cars will be connected and organized in networks for sharing information that will be first transmitted and processed in the cloud facilities of car manufacturers or suppliers. Such a system is called an Intelligent Transport System (ITS) and is aimed at using information and communication technologies of transport infrastructures to improve safety, reliability, efficiency and quality for all travels by car [1].

In this context, Continental Digital Service France (CDSF) and LAAS-CNRS started the eHorizon project (2017-2021) for addressing the research and technological issues of such ITS. This paper then deals with presenting the global communication architecture for the ITS, as well as its requirements in terms of communications, including reliability and QoS parameters. It is especially required for this system to be highly flexible to adapt to any situation and use case, and to provide very low latency. For this purpose, the ITS system relies on the new and original Software Defined Networks (SDN) concept in the context of ITS. This concept is detailed in the paper, and its benefits are evaluated on some very representative scenarios. This paper is organized as follows. Section II gives a detailed description of related works. Section III presents the global communication architecture of an intelligent transport system and the major challenge of ITS communications. Section VI presents some ITS Services and their network requirements. Section V describes the proposed architecture. Section VI presents the proposed use cases. Section VII presents the experimental results. Finally, section VIII concludes this paper.

## II. RELATED WORK

In the last few years, with the emergence of new ITS services, vehicular networks are attracting more attention. Various scientific research organizations, industries and standardisation bodies are interested in improving them by proposing new architectures and mechanisms in order to effectively support these new services.

To that end, "I. Ku" and al. [9] propose to apply SDN to VANET networks in order to control the inter-vehicle ad-hoc communications and vehicle to RSU communications. A routing algorithm that exploits the global view of the SDN controller is proposed and performance tests show better results compared to traditional VANET routing protocols, which motivates the application of the SDN paradigm in that context. To address the problem of resiliency raised by the use of a single controller, a fallback mechanism based on a local controller embedded in the vehicle is also proposed to run traditional routing protocols in case of connectivity loss with the controller. However, this approach introduces an additional user cost. Based on the architecture of [9] and the global vision provided by the SDN controller, other routing algorithms were also proposed in [10, 11] and compared, by simulation, to traditional VANET routing protocols to show the performance benefits of SDN even when the programmable nodes are moving vehicles. In the same direction, "Y. C. Liu" and al. [12] propose a Software Defined Network (SDN) architecture for GeoBroadcast in VANETs in which the RSU entities are openflow enabled, and connected via openflows switches, all under the control of an SDN controller. Performance tests show that with programmable RSUs, better performance is achieved in comparison to the GeoNetworking protocol [13] used in traditional ITS architectures. "N. B. Truong" and al. [14] explore the use of Fog computing in an SDN-VANET architecture in order to effectively support low latency services. In [15], a Decentralized Software-Defined VANET Architecture is proposed where the SDN controller is distributed to address the scalability issue inherent to very dense environment. Results show that, control plane distribution improves network scalability, in addition to keep better delivery packets delays.

Our proposition differs from previous works in two main directions, Firstly, by applying SDN to the global ITS communication architecture including not only ad-hoc or RSU networks, but also the cellular network in a hierarchical manner in order to enforce a scalable network. This paves the way to the development of new network control mechanisms that take advantage of the ability to simultaneously control all available network resources. Secondly, we leverage the data collected by cloud platforms of the ITS service providers [16] to devise wise, proactive and effective network control algorithms that are aware of the environment and the context in which vehicles evolve (density and speed of vehicles, weather conditions,...) and may evolve in the near future. Indeed, with these data, an estimate of the network topology, network and node loads can be predicted and potential changes in network conditions can be anticipated and treated proactively.

The following table summarizes our positioning w.r.t related work

| Reference | Programmable components of the architecture that can be used for routing | | | SDN Controller | Using Data present in the cloud |
|---|---|---|---|---|---|
| | Vehicle | RSU | BS | | |
| [9] | √ | √ | | centralized | |
| [10] | √ | | | centralized | |
| [11] | √ | | | centralized | |
| [12] | √ | √ | | centralized | |
| [15] | √ | √ | | hierarchical | |
| [14] | √ | √ | √ | hierarchical | |
| Our proposition | √ | √ | √ | hierarchical | √ |

TABLE I. POSITIONING W.R.T RELATED WORK

III. THE GLOBAL COMMUNICATION ARCHITECTURE OF AN INTELLIGENT TRANSPORT SYSTEM

In this section, we describe the global communication architecture of an ITS. We first present the components of this architecture and their interactions, the technologies and standards used. Then, we discuss the different communication challenges that arise in such system.

A. Main Components

The global architecture consists of three main parts as shown in Fig.1: The vehicle, the network infrastructure managed by an Internet Service Provider (ISP), and the cloud platform controlled by an ITS service provider:

• **Vehicle:** A vehicle is equipped with a set of sensors and systems (GPS, Radar, Lidar, Advanced Driver-Assistance System (ADAS) camera, etc.) enabling it to collect several information about its environment (position, speed, neighboring vehicles, temperature, etc.). Depending on its location, it can be reached, as shown in Figure 1, only by a Road Side Unit (RSU) (vehicle A), or only by a Base Station (vehicle B), or both (vehicle C), or it may be out of any network coverage (vehicle D). A vehicle can be equipped with several interfaces allowing it to interact with the various components of the system: (1) a 3/4G interface enabling it to benefit from different functionalities offered by the cellular network (Internet access, communication with other parts of the system (Vehicles, Cloud, etc.)), (2) a Dedicated Short Range Communication (DSRC) Interface enabling it to communicate with the RSU entities as well as other vehicles equipped with the same interface, and (3) a short range wireless Interface (e.g. Bluetooth) allowing it to communicate with the connected objects that surround it, as well as with the different User Equipments (UE) handled by the pedestrians, as illustrated in Figure 1. A vehicle acts not only as an end node, but also as a router to transmit information to other vehicles.

• **Network Infrastructure:** The network infrastructure is composed mainly of two parts : RSU and cellular network:

• *Road Side Unit:* A RSU entity represents one of the dedicated components for an ITS system. It may be implemented in a base station, or in a dedicated stationary entity installed along the road. It is mainly equipped with a DSRC interface, with which, it can communicate with any component equipped with the same interface (vehicle, RSU, etc.). Its communication range depends on the environment and the technology used. For example an RSU entity that supports the DSRC standard can have a communication range of 300 m in urban environnements, and a communication range up to 1km in rural environnements. The Road Side Units may be interconnected via a wired or wireless medium, and they can not only provide a local service but also a cloud service and/or Internet access to the different vehicles.

• *Cellular Network:* The cellular network represents one of the main technologies that may support the different vehicular communications. It has a very high network capacity enabling it to support applications requiring high throughput/bandwidth demands.

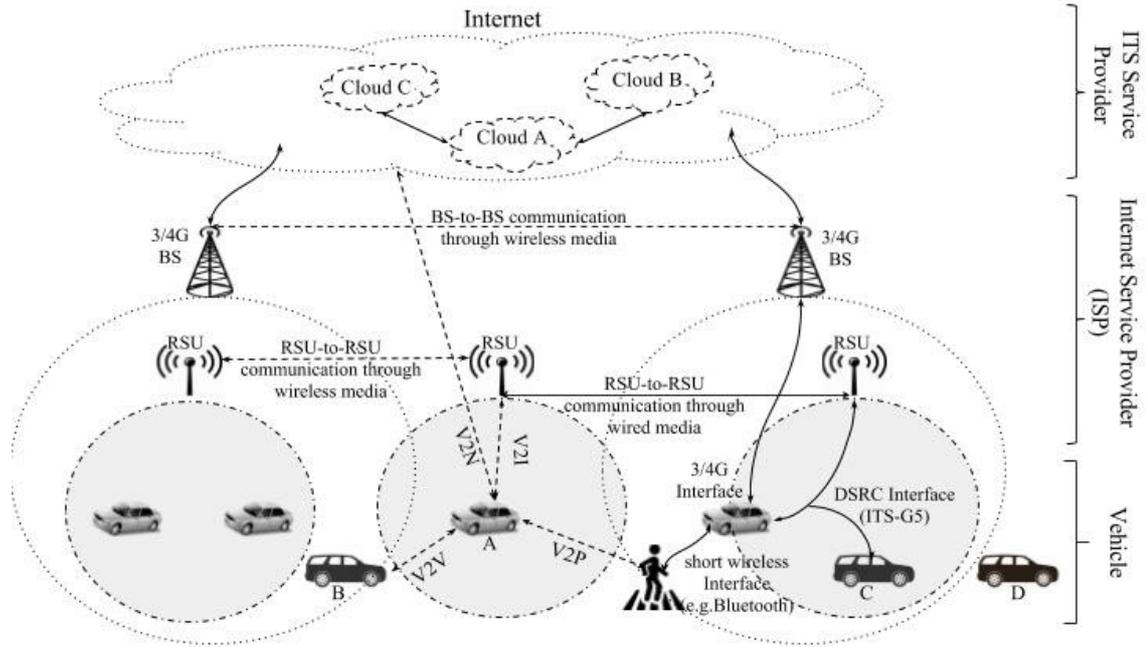

Fig. 1. GLOBAL COMMUNICATION ARCHITECTURE OF IN INTELLIGENT TRANSPORT SYSTEM

It has a very high network capacity enabling it to support applications requiring high throughput/bandwidth demands. Moreover, it is characterized by a wider communication range, which allows a base station to maintain connectivity with a network node (vehicle) as long as possible, thereby limiting handover operations. In addition, It offers Multicast/ Broadcast transmission services (MBMS/eMBMS) and D2D communication technology, which can be used extensively in an ITS system.

- **Cloud/Fog computing:** The cloud computing is the intelligent part of the system. It has a high storage and processing capabilities to massively collect data, and process them to provide customized ITS services to different vehicles, whereas Fog computing represents a distributed data-center whose computing devices are closer to the end users in order to provide real-time services that require a very low latency.

*B. Communication Types*

A vehicle can interact with its environment through various types of communication, as presented in Figure 1, and specified in [2]:

- **V2V** (Vehicle-to-Vehicle): A type of communication, in which both communicating parties are UEs (vehicles) using V2V applications.

- **V2P** (Vehicle-to-Pedestrian): A type of communication, in which both communicating parties are UEs (vehicle, pedestrian) using V2P applications.

- **V2I** (Vehicle-to-Infrastructure): A type of communication, in which one part is a UE (vehicle) and the other part is an RSU entity, both using V2I applications.

- **V2N** (Vehicle-to-Network): A type of communication, in which one part is a UE (vehicle) and the other part is a serving entity, both using V2N applications.

*C. Communication Challenges*

The density and the speed of the vehicles are the major factors affecting the quality of the vehicular communications. In high density networks, vehicles must efficiently share available network resources in order to avoid congestion problems, which is a challenging task. Besides, the high speed of the vehicles complicates the maintenance of the communication between the nodes. This situation becomes more complicated when the vehicles move in opposite directions. Table II shows the variation of the speed and the density of vehicles per environment (urban, suburban and highway) as described in [3]. It is noticed that when the density of the vehicles decreases, the speed of the vehicles and the range of communication increase.

| Scenario | Vehicle density (vehicles/km2) | Relative Velocity (Km/h) | Communication range (m) |
|---|---|---|---|
| Urban | 1000-3000 | 0-100 | 50-100 |
| Suburban | 500-1000 | 0-200 | 100-200 |
| Highway | 100-500 | 0-500 | 200-1000 |

TABLE II. CHARACTERISTICS OF THE DIFFERENT ENVIRONMENTS

IV. SERVICES AND THEIR REQUIREMENTS

In this section, we present some ITS services and their network requirements. We then discuss some communication technologies that may support these requirements.

ITS services can be classified into safety and non-safety services as specified in [4]:

- **Safety Services:** The main objective of these services is to improve the road safety by minimizing the number of accidents and reducing the possibilities of life loss. They require a very low latency and high reliability. Some safety services and their network traffic models are presented in Table IV.

- **Non Safety Services:** The main objective of these services is to improve the traffic efficiency (avoid traffic jams, optimize transport times and gas consumption, etc.) and to provide to the vehicle's users some services (infotainment, Internet access, etc.) that enhance their experience. These services have no stringent demands on latency and reliability compared to safety services. Some non safety services and their network traffic models are presented in Table IV.

The network traffic models are derived from a requirement analysis of these services as specified in [4]. A traffic is defined by a behavior (periodic or non-periodic), a transmission mode (unicast, broadcast), a maximum latency, a minimum frequency for the periodic messages, and a level of transmission reliability. Other elements can be considered such as the required security mechanisms. Table III presents the identified network traffic models. Table IV summarizes the models that characterize the traffic generated by each service.

| Model 1 | periodic, broadcast, maximum latency=100ms, minimum frequency=10 Hz, high reliability requirements |
|---|---|
| Model 2 | non-periodic, unicast, maximum latency=100ms, minimum frequency=10 Hz, high reliability requirements |
| Model 3 | periodic, broadcast, maximum latency=500ms, minimum frequency=2 Hz, low reliability requirements |
| Model 4 | non-periodic, unicast, maximum latency=500ms, minimum frequency=2 Hz, low reliability requirements |

TABLE III. NETWORK TRAFFIC MODELS

| ITS Services | Use Case | Usage | M1 | M2 | M3 | M4 |
|---|---|---|---|---|---|---|
| Safety Services | Co-operative forward collision warning | Avoid longitudinal collision | √ | √ | | |
| | Emergency vehicle warning | Reduce emergency vehicle's intervention time | | √ | | |
| | Wrong way driving warning | Limit as much as possible frontal collisions | √ | | | |
| Non Safety Services | Traffic information and recommended itinerary | Traffic information and regulation | | | √ | |
| | Automatic access control/ parking access | Facilitate vehicle access to controlled areas | | | √ | √ |
| | Remote diagnosis and just in time repair notification | Reduce the risk of vehicle failure | | | √ | √ |

TABLE IV. ITS SERVICES AND THEIR NETWORK TRAFFIC MODELS

Two main technologies are considered to support these requirements, DSRC and cellular technologies (LTE, Long Term Evolution [18]), However, these technologies have some limitations, making them unable to efficiently support these services. The short range communication of DSRC entities (e.g. RSU) limits their ability to offer the services that require continuous data dissemination along the road (e.g. data streaming, online gaming, Internet access), especially when the vehicles move at high speed. In addition, the CSMA/CA (Carrier Sense Multiple Access with Collision Avoidance) [19] technique used by the DSRC standards to avoid collision introduces a significant access delay to the channel, which causes scalability problems especially in dense environments. On the other hand, cellular technologies are not suitable to support V2V communications which require a very low latency, due to the centralized architecture of cellular networks. The studies in [5] and [6] show that the performance of each technology drops once the speed and the density increase. Consequently, A new architecture with new mechanisms is required to efficiently support these new services.

## V. AN SDN HYBRID ARCHITECTURE

In this section, we describe the proposed SDN architecture. We first present the SDN concept, then the advantages of applying it to the global architecture of an ITS system.

### A. Software Defined Network

SDN is an emerging network paradigm which advocates the idea of taking control plane functions out of network forwarding devices and relocating them on remote external computing machines called SDN controllers. The network intelligence and state are logically centralized [7]. The SDN controller communicates with the different network nodes using a southbound interface protocol, i.e. the widely used OpenFlow standard [8], while applications explicit their requirements to the SDN controller using Northbound Interface (API), as presented in Figure 2. In this architecture, the network nodes forward packets according to the rules installed on network devices by the SDN controller in a proactive or reactive manner.

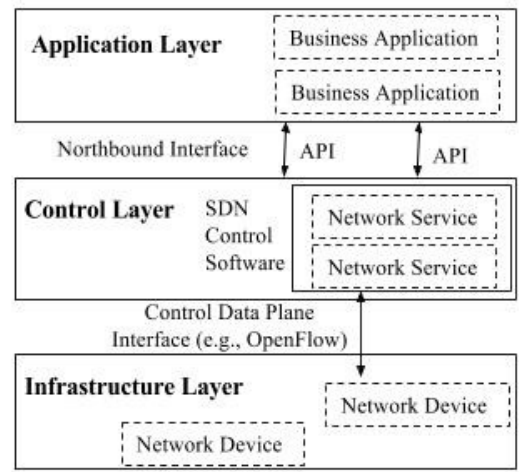

Fig. 2. SOFTWARE DEFINED NETWORK ARCHITECTURE [7]

### B. Benefits of Our Approach

In our approach, we apply SDN to the global architecture of an ITS, including, not only the ad hoc network as proposed in [9], but also the RSU and cellular networks. This approach responds to the limitations of current architectures, by opening the road to the development of novel network control algorithms that take advantage of: (1) a vision of the state of all three above cited communication networks; (2) the ability to jointly control these networks; and (3) the knowledge of the environment in which vehicles evolve, which is derived from the data present in the cloud. For example, vehicles status information (position, direction, speed) can be used to predict the number of vehicles that will be present in a given region at a given time, allowing the estimation of the potential network load of a routing node (BS/RSU). Moreover, the dynamic nature of vehicular networks requires an adaptable network, SDN brings this flexibility to dynamically program the network according to network conditions. The SDN controller, which is a new component, is added to the architecture, as illustrated in Figure 3. Typically, We consider three controllers: one to manage the cellular network, another to manage the RSU-based network and a last one to coordinate between the different controllers. The main controller builds a global view of the communication infrastructure using the information sent by the controller of each network coupled with the data present in the cloud. It sends to each controller the global rules which describe the general behavior of the network, while the BS and RSU controllers define the specific rules to be installed in each network device. The communication between the SDN controllers is done using a specific interface known as East-West Interface e.g. AMQP, while the communication between the SDN controllers and the cloud is performed through specific APIs.

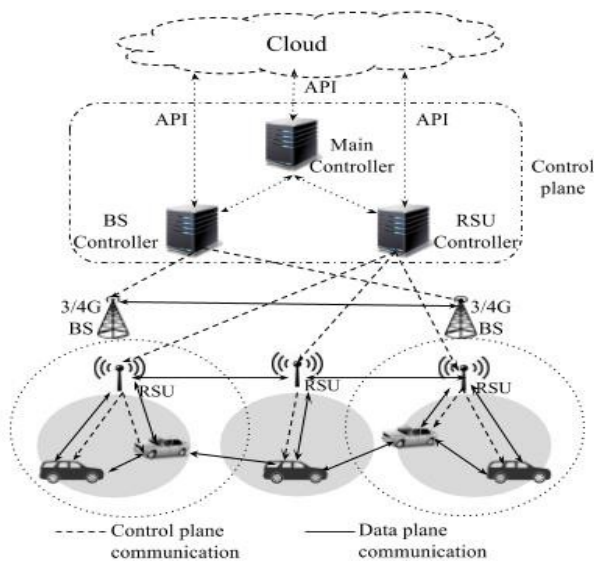

Fig. 3. THE PROPOSED SDN ARCHITECTURE

Among the opportunities brought by this architecture:

- **QoS aware routing with potential environmental inputs in a multi homed context:** The SDN controller provides the best routing path according to the services requirements through efficient routing algorithms that are aware of the QoS requirements of each ITS services and the environment in which vehicles evolve, and which take advantage of the presence of several networks.

- **Mobility Management:** The global vision of the network allows the SDN controller to provide a better coordination of handover operations, moreover, the collected data present in the cloud allows it to predict the mobility of surrounding vehicles in order to anticipate some control operations.

- **Enhanced QoS Management:** The QoS management can be improved thanks to the fine grained as well as on-line programming capabilities offered by SDN. Efficient and dynamic QoS support can be achieved. Joint control algorithms (routing, topology control, etc.) can be developed to that end.

- **Network Load Balancing and flow splitting**.

## VI. USE CASES

In this section, we present some very representative use cases in order to show the benefits of our approach.

### A. Cooperative collision avoidance

We consider the "cooperative collision avoidance" service as one of the main safety services of "self-driving". The primary goal of this service, as its name suggests, is to help vehicles avoid crashes. The Vehicles continuously exchange information about their trajectories and status (position, velocity, direction). Therefore, each vehicle uses these information to compute the optimal collision avoidance actions and apply them in a cooperative manner. As specified in [3] and [4], the communications between vehicles must be made within a maximum latency of 100 ms, and shall not fail with a probability higher than $10^{-5}$ (the service tolerates the loss of one packet among $10^5$ packets sent) which represents a significant challenge for the network. Let's refer the traffic of this service as (A1) and let's consider that the vehicles simultaneously execute other services (A2, A3) which have no stringent requirements on latency and reliability compared to A1.

In legacy vehicular ad-hoc networks [17], a priority based medium access scheme is defined with four priority levels named Access Categories (AC). Traffic is directed to the appropriate AC according to its priority and is statistically given priority in comparison to lower priority traffic originating from other véhicles. Despite this priority scheme, with no admission control (which is hard to set up in an ad-hoc context), there is no guarantee that the highest priority traffic (A1) will receive its requested network performance, particularly when entering a crowded area with high priority traffic from multiple vehicles contending for transmission.

With our proposed SDN based architecture, thanks to the centralized global view of the network, network resource allocations can be envisioned in order to provide the required QoS to each service, with the added ability to keep up with varying QoS requirements (e.g. referring to the figure below, a channel has been allocated dynamically to A1, while A2 and A3 share the same channel). Thanks to the multihomed nature of vehicles, these allocations can simultaneously apply to the different active connecting networks (cellular, RSU, ..) which paves the way to effective QoS provisioning algorithms.

There is another advantage that traditional architectures cannot achieve compared to our proposed architecture: the ability to reconfigure the network according to the changes of the network conditions (link disruption, failure of a network node, interference because of node density). We consider the case where the quality of wireless transmissions is degraded due to weather changes, which directly affects the transmission reliability and hence the functioning of the service.

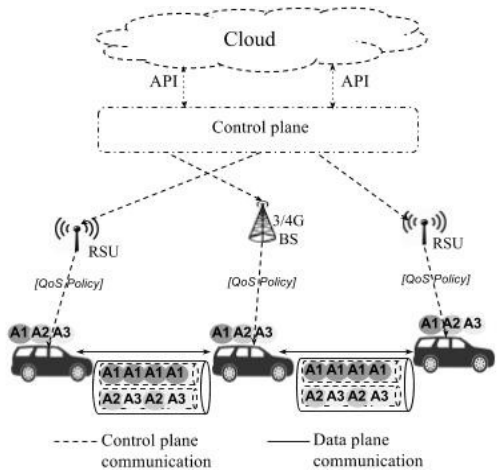

FIG. 4. COOPERATIVE COLLISION AVOIDANCE-SCENARIO 1

In traditional architectures, the vehicle can detect these changes in a reactive manner by continuously monitoring the quality of wireless links (SNR, BER). However the decisions that it can take are predefined and limited to its local knowledge. For example, it can duplicate the critical traffic on two different channels, or change the transmission parameters (modulation technique...), which may not be relevant in some cases.

In our proposed architecture, the SDN controller can detect proactively these changes using data from the cloud. For example, the weather forecast in a given region can be exploited by network control applications to predict forthcoming changes in transmission conditions (wireless link quality), and take the appropriate actions to overcome this problem as, for example, duplicating the A1 traffic on two different paths in order to increase the packet delivery probability and consequently, provide A1 with the expected transmission reliability (see figure 5).

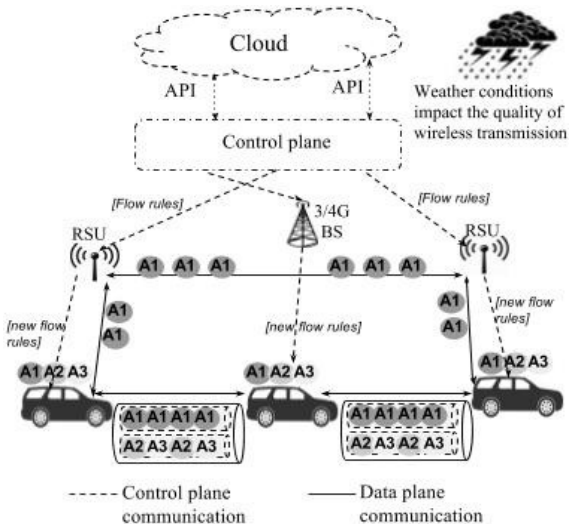

Fig. 5. COOPERATIVE COLLISION AVOIDANCE-SCENARIO 2

### B. Bird's eye view

Among the ITS services proposed by the standardization bodies, we find those based on "Data Streaming", for example, road traffic monitoring applications, multimedia applications. In our case, we consider the safety service " Bird's eye view" allowing the vehicles equipped with sensors such as cameras, radars, lidars to share their views with the neighboring vehicles. The vehicles can use this data (maybe fusioned with other data provided by other sources) in order to identify the pedestrians, free places and to better plan their future trajectories. This service requires high data rate (up to 40 Mbit/s), low latency (<50ms), and ubiquitous connectivity, which represents a great challenge in this very dynamic environment. We consider the scenario presented in the figure below, where vehicle V1 attempts to send the video captured by its camera to the vehicles V2, V3, V4. We assume that the different vehicles are equipped with DSRC and LTE interface.

When streaming Video Data to others vehicles, vehicle V1 must choose which technology to use. In traditional architectures, V1 will choose based on its local knowledge, using, for example, the link quality information (RSSI, SNR,) offered by each node (RSU, BS), however this choice may be ineffective, when the selected node is the most overloaded.

With our proposed architecture, we exploit the global vision of the state of each network, to make a more optimal choice according to the position of each vehicle and taking into account the network load of different nodes as well as the quality of various links. For example, the path V1-V2 is established using the 3/4G network, while the path V1-V3 is established using the RSU network, as shown in the figure below. In addition, in order to ensure ubiquitous connectivity, some vehicles mobility prediction algorithms, that exploit data present in the cloud, are used to predict, for example, the next RSU that will cover such a moving vehicle, which allows the SDN controller to install some control rules in a proactive manner.

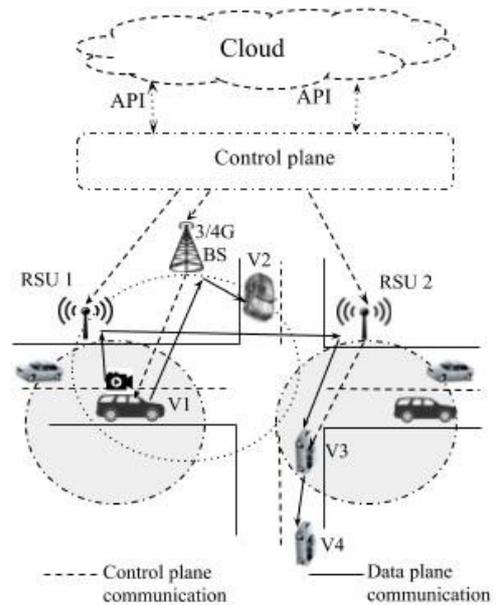

Fig. 6. BIRD'S EYE VIEW

## VII. EXPERIMENTAL RESULTS

The goal of the experimentation is to demonstrate how the global network view established at the controller combined and enriched with information brought from the cloud enables a wiser and more efficient control of network behaviour with, at the end, a service with enhanced performance provided to the user. To that end, we consider the "Bird's eye view" scenario presented in the previous section, and show through evaluations how the SDN controller can leverage its global view of current and forthcoming (from the cloud) network loads to guide the node in the selection of the point of attachment to the network with the best expected performance

### A. Simulation description

In order to simulate our scenario, we use the MiniNet-WiFi emulator [20], which is an extension of MiniNet [21] to emulate 802.11 wireless networks programmable via SDN. In an area of 2000 × 2000 m², we consider a network topology composed of 4 RSU entities, each with a communication range of 600 m, interconnected via wired connections; All are under the control of an SDN/Openflow controller.

The simulation plays out in two phases, a first one where the network has an average load, and a second one where the network is overloaded.

In the first phase, we use 10 vehicles running an udp client-server session using the Iperf tool [22] generating a given network traffic. The vehicle "car1" (server) attached to the RSU1, stream a video traffic that vehicles "car2" and "car3" (clients) covered respectively by "RSU2" and "RSU4" consume as shown in the Figure 7. Table V shows the characteristics of this traffic. The Vehicles covered by several RSU entities, select the network to which they attach according to the power of the received signal "RSSI".

| Traffic Type | UDP |
|---|---|
| Report Interval | 1 s |
| Buffer Size | 41 Mbytes |
| Bandwidth | 10.0 MBytes/s |
| Time | 60 s |
| Packet size | 1500 Bytes |
| Propagation Model | Log-Distance Propagation Loss Model (exp=3) |

TABLE V. TRAFFIC CHARACTERISTICS

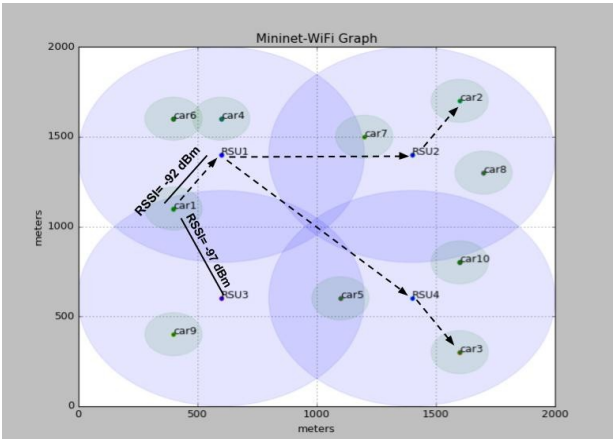

FIG. 7. PHASE 1 : NETWORK WITH AN AVERAGE LOAD

In the second phase, we add 5 vehicles in order to overload the RSU1 to which "car1" is associated (as shown in Figure 8). This network condition change will be anticipated by the SDN controller in order to apply some network control actions in a proactive way. In this scenario, the idea is to prompt vehicle "car1" to attach to RSU3. New flow rules will be installed in the RSU entities by the SDN controller to support this new flow.

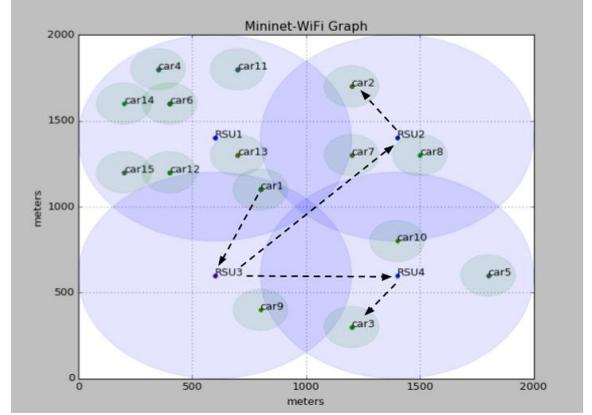

FIG. 8. PHASE 2 : NETWORK OVERLOADED

### B. Performance metrics

Two metrics are considered in our evaluations :
- *RTT* : Round Trip Time, is the time required for a packet to travel from a specific source to a specific destination and back again. This metric is measured using the ping utility.
- PDR : Packet Delivery Ratio, This metric represents the ratio of the delivered packets to the destination to those produced by the source node, which represents the reliability of the transmission, this metric that we measure using the Iperf tool.

### C. Simulation results

Figure 9 and 10 show the performance results for the communication between the "car1" and "car2" during the two simulation phases (average load and overloaded) and in both cases with (solid lines) and without (dotted lines) SDN controller presence. They respectively show the RTT and PDR as a function of time.

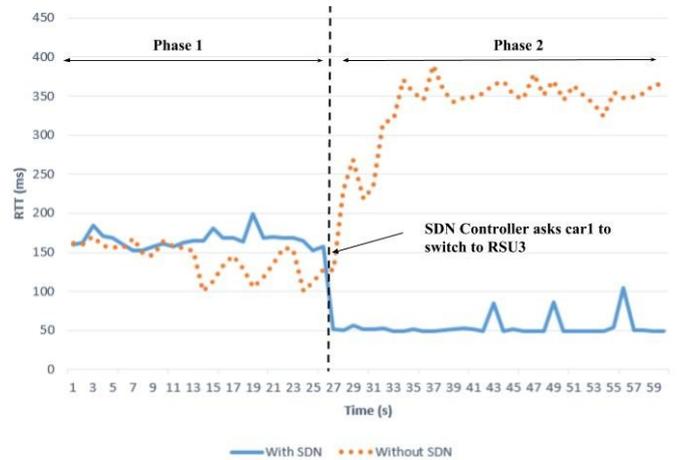

FIG. 9. RTT (ROUND TRIP TIME)

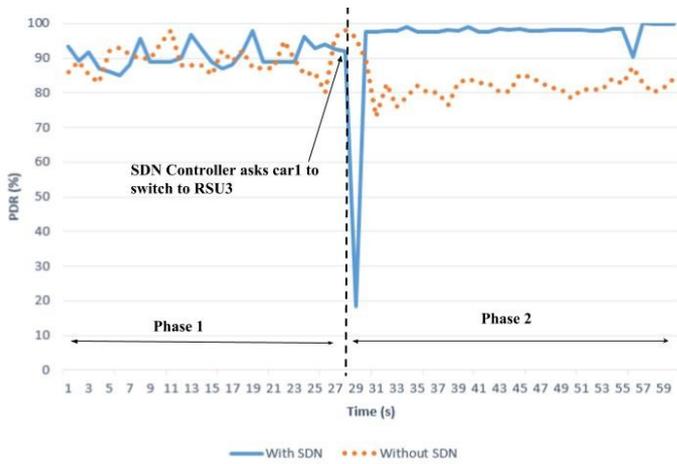

FIG. 10. PDR (PACKET DELIVERY RATIO)

We notice that during the second phase, when the RSU1 entity covering the vehicle "car1" is overloaded, the network performance degrades, the average RTT increases from 140 ms to 353 ms and the PDR decreases by 10%.

With the SDN controller which triggers the change of the RSU entity to which vehicle "car1" is attached, we remark that the network performances is improved, the PDR remains around 98% and the average RTT is decreased by 87 ms. However, this handover action has a cost in terms of network performance as we notice that the PDR drops to 20% during the RSU entity change.

## VIII. CONCLUSION AND FUTURE WORK

With the aim to improve the quality of service offered by vehicular networks, this paper aims at presenting a new architecture based on the SDN paradigm combining the RSU and the cellular networks in order to efficiently support the QoS requirements of the envisioned ITS services, combined with the data collected in the cloud, we argue that novel network control algorithms can be devised to improve the efficiency and QoS capabilities of vehicular networks.

We are now working on the development and evaluation of new network control functions (routing, QoS aware resource allocation, etc.) algorithms which take benefit of the proposed architecture.

ACKNOWLEDGMENT

This work is funded by Continental Digital Service France (CDSF) in the framework of the eHorizon project.